\documentclass[aps,pra,superscriptaddress,showpacs,twocolumn,10pt]{revtex4-1}
\usepackage[colorlinks=true,linkcolor=blue,urlcolor=blue,citecolor=blue]{hyperref}

\usepackage{dsfont}
\usepackage{xspace}
\usepackage{color}
\usepackage[usenames,dvipsnames]{xcolor}
\usepackage[colorlinks=true,linkcolor=blue,urlcolor=blue,citecolor=blue]{hyperref}
\usepackage[normalem]{ulem}

\usepackage{mathrsfs}
\usepackage{fullpage}
\usepackage{amssymb}
\usepackage{graphicx}
\usepackage{amsmath}
\usepackage[strict]{changepage}
\usepackage{framed}
\usepackage{amsthm}
\usepackage{bbm}
\usepackage{colortbl}

\def\bea{\begin{eqnarray}}
\def\eea{\end{eqnarray}}
\def\be{\begin{equation}}
\def\ee{\end{equation}}

\DeclareMathOperator{\tr}{\mbox{tr}}

\renewcommand{\d}{\ket{\downarrow}}
\renewcommand{\u}{\ket{\uparrow}}

\renewcommand{\be}{\hat b}
\newcommand{\bed}{\hat b^\dagger}

\def\bra#1{\mathinner{\langle{#1}|}}
\def\ket#1{\mathinner{|{#1}\rangle}}
\newcommand{\braket}[2]{\langle #1 | #2 \rangle}

\definecolor{dgreen}{rgb}{0.0, 0.5, 0.0}

\newcommand{\veck}{\mathbf k}
\newcommand{\vecR}{\mathbf R}

\newcommand{\vecq}{\mathbf q}

\newcommand{\vecr}{\mathbf r}

\begin{document}
\title{A mesoscopic Rydberg impurity in an atomic quantum gas}

\author{Richard Schmidt}
\affiliation{ITAMP, Harvard-Smithsonian Center for Astrophysics, Cambridge, MA 02138, USA}
\affiliation{Department of Physics, Harvard University, Cambridge MA 02138, USA}
\author{ H. R. Sadeghpour}
\affiliation{ITAMP, Harvard-Smithsonian Center for Astrophysics, Cambridge, MA 02138, USA}
\author{E. Demler}
\affiliation{Department of Physics, Harvard University, Cambridge MA 02138, USA}

\begin{abstract}

Giant impurity excitations with large binding energies are powerful probes for exploring new regimes of far out of equilibrium dynamics in few- and many-body quantum systems, as well as for in-situ observations of correlations. Motivated by recent experimental progress in spectroscopic studies of Rydberg excitations in ensembles of ultracold atoms, we develop a new theoretical approach for describing multiscale dynamics of Rydberg excitations in quantum Bose  gases. We find that the crossover from few- to many-body dynamics manifests in a dramatic change in spectral profile from resolved molecular lines to broad Gaussian distributions representing a superpolaronic state in which many atoms bind to the Rydberg impurity. We discuss signatures of this crossover in the temperature  and density dependence of the spectra. 
\end{abstract}
\maketitle

\textit{Introduction.---} Dynamics of impurities in quantum systems presents a particular challenge in quantum many-body physics. Their many-body character is manifest in the orthogonality catastrophe, where the change of the internal state of one impurity leads to a dramatic reconstruction of the entire many-body system \cite{ande1967}. Yet they do not allow a hydrodynamic description commonly used for other non-equilibrium processes, where the evolution of the many-body state can be reduced to the dynamics of a few collective degrees of freedom \cite{haug2008,schafer2009}. Likewise, a description based on Boltzmann rate equations is insufficient since scattering on an impurity leads to coherence and entanglement \cite{rosch1999,beenakker2007}.

Until recently the main motivation for studying impurity dynamics came from the study of electron transport in mesoscopic and nanoelectronic systems \cite{lee1985,van2002}.   Theoretical work focused almost exclusively on fermionic systems \cite{prok2008,chevy2010rev} discussing such spectacular phenomena as the non-equilibrium Kondo effect \cite{hewson1997,Goldhaber1998} and the orthogonality catastrophe \cite{nozieres69}. In contrast, studies of impurity problems in bosonic systems were limited to equilibrium properties such as the polaron binding energy and effective mass \cite{kala2006,devreese2009,Shchadilova2014}. However, recent efforts to probe non-equilibrium polaronic dynamics in bosonic gases have motivated a renewed theoretical interest in its richness \cite{Rath2013,Scelle2013,Shashi2014}.

\begin{figure}[t]
        \centering
        \includegraphics[width=0.9\linewidth]{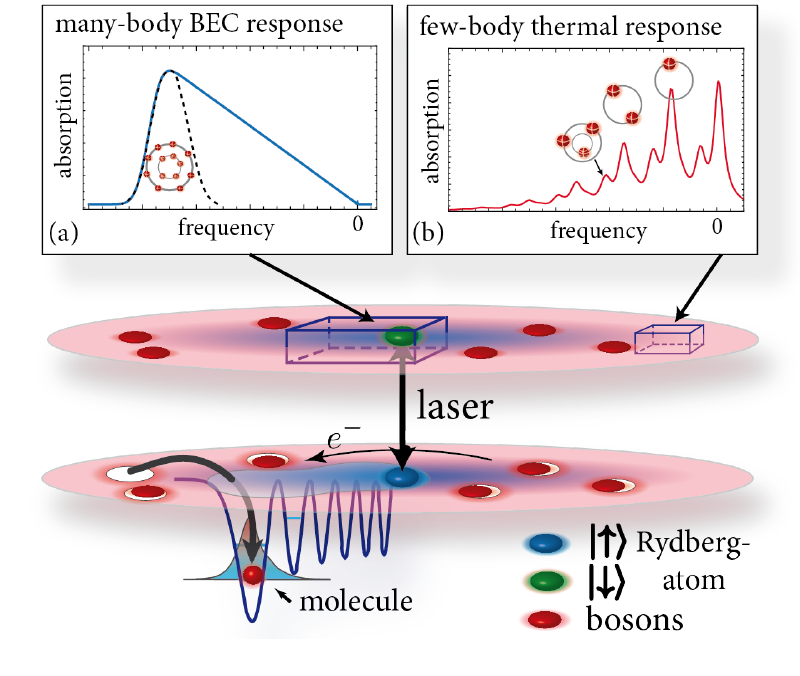}
             \vspace{-3mm}
  \caption{ \scriptsize When a Rydberg impurity is excited in a Bose gas (red spheres), a molecular potential supporting ultra long-range vibrational states is formed. When the excitation is performed in the thermal region of a trapped atomic gas, these state are observed as few-body absorption lines (dimer, trimer, tetramer, etc.), in which several particles bind to the impurity. The occupation of vibrational modes results in a shell structure of the collective bound state as illustrated in (b). If the excitation is created deep in a condensed gas, the spectrum becomes a normal distribution [dashed black curve in (a)] representing a metastable superpolaronic state of a mesoscopically large number of atoms. In reality, the laser excitation samples a range of densities (wireframes) leading to the observation of a density averaged spectrum [blue curve in (a)].}
        \label{fig.illustration} 
        \vspace{-6mm}
\end{figure}

The realization of highly-excited Rydberg impurities in ultracold quantum gases \cite{BalewskiNature13,gaj2014,nguyen2015} presents a new frontier where microscopic atomic physics  meets  condensed matter and mesoscopic physics, which goes beyond the idea of quantum simulations of strongly correlated states \cite{bloch2012}. At high principal quantum numbers, the Rydberg blockade mechanism ensures that only one excitation is possible in the gas \cite{Urban2009} promoting the Rydberg atom to a novel realization of an impurity problem. At the same time Rydberg excitations in ultracold gases were predicted to form exotic molecular bound states; dubbed trilobite molecules \cite{greene2000,bendkowsky2009}. Their mesocopic size combined with large binding energies, present a  multiscale challenge in impurity physics. Here the ultracold bosonic gas is dynamically probed on scales ranging from the deep binding energies  to the time scales of typical bath excitations, requiring the confluence of powerful theoretical tools, from atomic few-body and condensed matter many-body physics.

In this work, we address the challenge of non-equilibrium dynamics of Bose polarons arising from the creation of  Rydberg impurities and demonstrate how their dynamics manifests itself in the spectrum. Our approach employs microscopic molecular potentials calculated from  Rydberg wave functions \cite{Marinescu94}. Using exact bound and scattering solutions, we solve the full time evolution of the many-body density matrix, based on a novel functional determinant approach to bosonic systems. We show in detail how few-body molecular lines at low densities evolve into a giant superpolaronic state at higher densities. This dressed impurity state is comprised of a mesoscopic number of bound atoms and exhibits a shell structure in real space which can be probed in  experiments, see Fig.~\ref{fig.illustration}.

\textit{Model for Rydberg impurities.---} We investigate a single Rydberg impurity immersed in a Bose gas of ultracold atoms, localized in space at arbitrary temperature $T$ and particle density $\rho$.  The Rydberg atom is initially in its electronic ground state which we denote by $\d$. Using a two-photon laser transition the atom is transferred to a highly excited Rydberg state $\u$.  When driven into the $\u=\ket{ns}$ state,  the Rydberg electron of principal quantum number $n$, is in an orbital $\psi_e(\vecr)$ with an extent on the order of the interparticle distance $\rho^{-1/3}$ in the Bose gas. The scattering of the Rydberg electron from the surrounding ground state atoms is described by the Fermi pseudopotential method \cite{fermi1934}. This approach, based on the large separation of time scales for the electronic and interatomic motion, leads to a potential which scales with the Rydberg-electron perturber-atom scattering length $a_e$.   A host atom situated at a distance $\vecr$ from the impurity ion core at $\vecR= 0$ is subject to the potential \cite{gaj2014}
\begin{equation}\label{RydbergPotential}
V_\text{Ryd}(\vecr)=\frac{2\pi \hbar a_e}{m_e}|\psi_e(\vecr)|^2.
\end{equation}
This Born-Oppenheimer potential, calculated from Rydberg wave functions $\psi_e$ \cite{Marinescu94} has been shown to support bound  vibrational states \cite{greene2000}, when $a_e<0$; for an illustration cf.~Fig.~\ref{fig.illustration}. 

The main knob for the control of interactions in quantum gases has universally been the zero-momentum scattering length which can be manipulated by magnetic or optical Fano-Feshbach spectroscopy \cite{chin2010}. Here, in Rydberg impurity excitations, it is the variation of $|\psi_e(\vecr)|^2$  which brings in the possibility to probe the quantum gas in a completely novel way, i.e.~by changing the principal quantum number of the Rydberg atom.

So far, theoretical analysis of Rydberg molecule excitations \cite{Liu2006,Bendkowsky2010,DeSalvo2015,schlagmueller2015} have relied on few-body atomic methods which address the energies of individual states but that by construction cannot account for  many-body quantum dynamics.  In contrast, to describe  the physics on all length scales, arbitrary temperature and density, one must rigorously include the full quantum statistics, encoded in the many-body density matrix. This applies in particular to Rydberg excitations of a large principal number where the ultracold medium is probed on mesoscopic length scales and  standard many-body techniques such as mean-field or variational approaches fail due to the multiscale nature and the non-perturbative character of the underlying microscopic bound state physics. 

To this end, we study the dynamics of the Rydberg impurity governed by the many-body Hamiltonian 
\begin{eqnarray}\label{Hamiltonian}
\hat H&=&\sum_\veck \epsilon_\veck \bed_\veck \be_\veck+\sum_{\veck\vecq}V(\vecq)\bed_{\veck+\vecq}\be_\veck \ket{\uparrow}\bra{\uparrow}
\end{eqnarray}
where $\bed_\veck$, $\be_\veck$ represent the creation and annihilation operators of the bath bosons of mass $m$ with dispersion relation $\epsilon_\veck=\veck^2/2m$ and $V(\vecq)$ is given by the Fourier transform of Eq.~\eqref{RydbergPotential} seen by the atoms in the medium only if the Rydberg atom is in the excited state $\ket{\uparrow}$. The many-body Fock space is constructed from single-particle orbitals of the bosons. Those orbitals are obtained  from the bound and continuum eigensolutions of the Schroedinger equation for a localized Rydberg impurity with infinite mass. The scattering length in the pseudopotential in Eq.~\eqref{RydbergPotential} is rescaled to reproduce the bound molecular dimer energies. Since the formation of Rydberg molecules takes place on time scales governed by their deep binding energies, the Rydberg ion recoil can be ignored. Since this time scale is also short compared to the collective excitations of the bath, the inter-boson interaction can be neglected as well.

\textit{Rydberg impurity spectra from a quantum quench problem.---} We compute the time-dependent overlap 
\begin{equation}\label{overlap}
S(t)=\tr [ e^{i\hat H_0 t}e^{-i\hat H t} \hat \rho ]
\end{equation}
where $\hat \rho$ is the density matrix representing the initial state of the system, $\hat H$ is given by Eq.~\eqref{Hamiltonian}, and $\hat H_0$ is the Hamiltonian in the absence of the impurity. The expression~\eqref{overlap} describes the many-body dephasing dynamics following the sudden quench of the Rydberg potential. It can be directly measured in real time using Ramsey spectroscopy \cite{goold2011,knap2012,Cetinainprep,Schmidtprep}. The Fourier transform of Eq.~\eqref{overlap} yields the two-photon absorption spectrum $A(\omega)=2\,\text{Re}\int_0^\infty dt e^{i\omega t}S(t)$ \cite{Mahan90,knap2012}.

\begin{figure}[t]
        \centering
        \includegraphics[width=\linewidth]{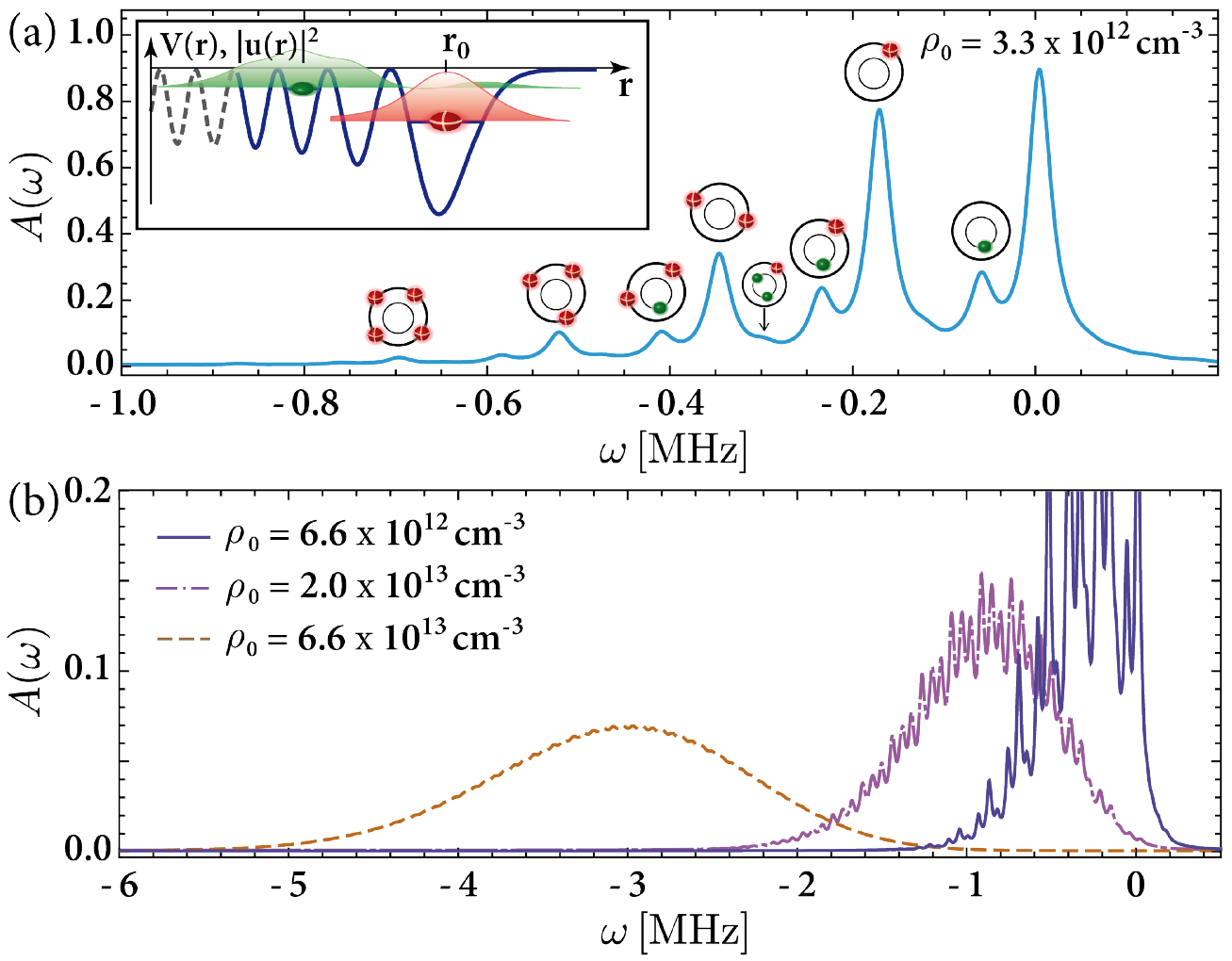}
        \caption{Absorption spectrum at $T=0$ for a Rydberg excitation to $^{87}\text{Rb}(71s)$ (excitation duration $70\mu$s). (a) Low density $\rho_0=3.3\times 10^{12}\text{cm}^{-3}$  at position of the Rydberg atoms. The circles with spheres above the peaks represent atoms bound in the different bound state orbitals and their spatial occupation probability $|u(r)|^2$ for the s-wave single-particle states is shown in the inset.  (b) Crossover to the many-body regime with increasing densities.
}
\vspace{0mm}
        \label{fig71s} 
\end{figure}   

\textit{Zero temperature description.---} We focus first on a Bose gas at zero temperature. The Fock state representing the macroscopic occupation of bosons in a BEC is given by  $\ket{\Psi_0}=1/\sqrt{N!} (\be^\dagger_0)^N\ket{\text{vac}}$ where $N$ is the boson particle number.  The time-dependent overlap  \eqref{overlap} is then evaluated with respect to the density matrix $\hat \rho_{\text{BEC}} = \ket{\Psi_0}\bra{\Psi_0}$ and gives
\begin{equation}\label{ZeroTS}
S_{N}(t)=\left(\sum_{\alpha} |\braket{\alpha}{s}|^2 e^{i(\epsilon_s-\omega_\alpha)t}\right)^{N},
\end{equation}
where the collective index $\alpha=(n,l,m)$ labels the interacting single particle states $\ket{\alpha}$ with quantum number $n$, angular momentum $l$ of projection $m$ and energy $\omega_\alpha$. The lowest non-interacting scattering state is denoted as  $\ket{s}=\be^\dagger_0\ket{\text{vac}}$ with energy $\epsilon_s$. 

We  demonstrate here the utility of our method for the specific example of  excitations in a $^{87}\text{Rb}$ gas, as recently realized in experiments \cite{bendkowsky2009,gaj2014,nguyen2015}. We focus on the excitation into a state of high principal quantum number Rb(71s). Vibrational bound and scattering states up to large angular momenta are calculated in a spherical box of radius $10\,\mu$m \cite{sup}. Furthermore we incorporate the finite length of excitation pulses in the calculation in Eq.~\eqref{overlap}. Similarly the finite lifetime of Rydberg excitations can  be included. The dominant contributions to the dynamics come from the few lowest vibrational single-particle bound states.   Hence it suffices to include the potential for the few outer most wells of the potential [cf.~inset in Fig.~\ref{fig71s}(a)]. 
  
  In Fig.~\ref{fig71s} we show the absorption spectrum for a Rydberg excitation in the BEC phase at T=0, for different densities $\rho_0=\rho(\vecr=0)$. The low density response is shown in Fig.~\ref{fig71s}(a). Here the interparticle spacing exceeds the range of the Rydberg potential, $r_0\approx8650 \, \text{a$_0$}$. The physics is dominated by few-body interactions resulting in a series of molecular lines corresponding to  one, two or more medium atoms bound and localized inside the Rydberg orbit. These lines, which are reminiscent of the few-body molecular spectra recently observed in a thermal Rb gas \cite{bendkowsky2009,gaj2014}, are identified in Fig.~\ref{fig71s}(a) with a red sphere attached to the outermost Rydberg orbital for the fundamental tone dimer line, the associated trimer line with two red spheres, tetramer and pentamer lines, with three and four red spheres, respectively.

The Rydberg potential Eq.~\eqref{RydbergPotential} supports several vibrational bound states (for the s-wave channel with binding energies $\epsilon_{B}\approx 175\, \text{kHz}$ and $\epsilon_{B}'\approx 64\, \text{kHz}$). This leads to additional spectral features represented by a combination of green and red spheres, which are associated with exotic bound dimers, trimers, tetramers, etc. The spatial structure of the single-boson wavefunctions [see inset in Fig.~\ref{fig71s}(a)] reveals that  atoms in the first excited vibrational orbit are localized at smaller distances from the Rydberg ion. In consequence these complex Rydberg molecules exhibit a spatial shell structure reminiscent of the shell model in nuclear physics \cite{blatt2012}.  

As density is increased, the character of dynamics changes from few- to many-body. In this regime, shown in Fig.~\ref{fig71s}(b), we predict a crossover from resolved molecular lines with an asymmetric envelope (blue and purple line) to a continuous Gaussian distribution (brown line) with a peak which moves progressively toward larger detuning as the density is increased. This broad spectral response corresponds to the formation of states with large number of atoms bound to the Rydberg impurity. Its emergence is rooted in the stochastic occupation of densely packed many-body states absent in typical condensed matter and ultracold atom realizations of impurity problems \cite{schirotzek_observation_2009,Kohs2012,Latta2011}.

The predicted Gaussian profile can be understood from the spectral decomposition of $A(\omega)$, in excitations from the BEC ground state $\ket{\Psi_0}$ into interacting single-particle states $\ket{\alpha_i}$ \cite{sup}. The resulting multinomial distribution is dominated by a few states with favorable Franck-Condon factors $p_i=|\braket{\alpha_i}{0}|^{2}$. If only two interacting single-particle states, the deepest bound state $\ket{B}$ and the zero-energy scattering state $\ket{\tilde s}$ are considered, a binomial distribution results \cite{sup},
\begin{equation}\label{specbinom}
 A(\omega)=\sum_{m=1}^N  \binom {N}{m} p^m (1-p)^{N-m}\delta(\omega-m\epsilon_B).
 \end{equation}
Here $p=|\braket{B}{s}|^2$ is the probability to transfer a BEC atom from the single-particle state $\ket{s}$ into the bound state.  For large particle number $N$, this distribution evolves naturally into a Gaussian.  Of course, in the exact spectrum shown in Fig.~\ref{fig71s} additional interacting states contribute, leading to a multinominal spectral decomposition of the absorption intensity. This distribution also evolves into a Gaussian distribution when the number of particles becomes large. 

Our analysis shows that Rydberg impurities in an ultracold  gas allow for a quantum realization of the urn problem. Here the BEC reservoir represents the urn while the Rydberg impurity facilitates the random drawing process. We emphasize that the Gaussian distribution, which is rooted in the wave nature of the particles at zero temperature, emerges solely due to the canonical nature of atomic BEC and we predict it to be absent in a grand-canonical BEC such as recently realized using dye filled cavities \cite{Schmitt2014}.

\begin{figure}[t]
        \centering
        \includegraphics[width=\linewidth]{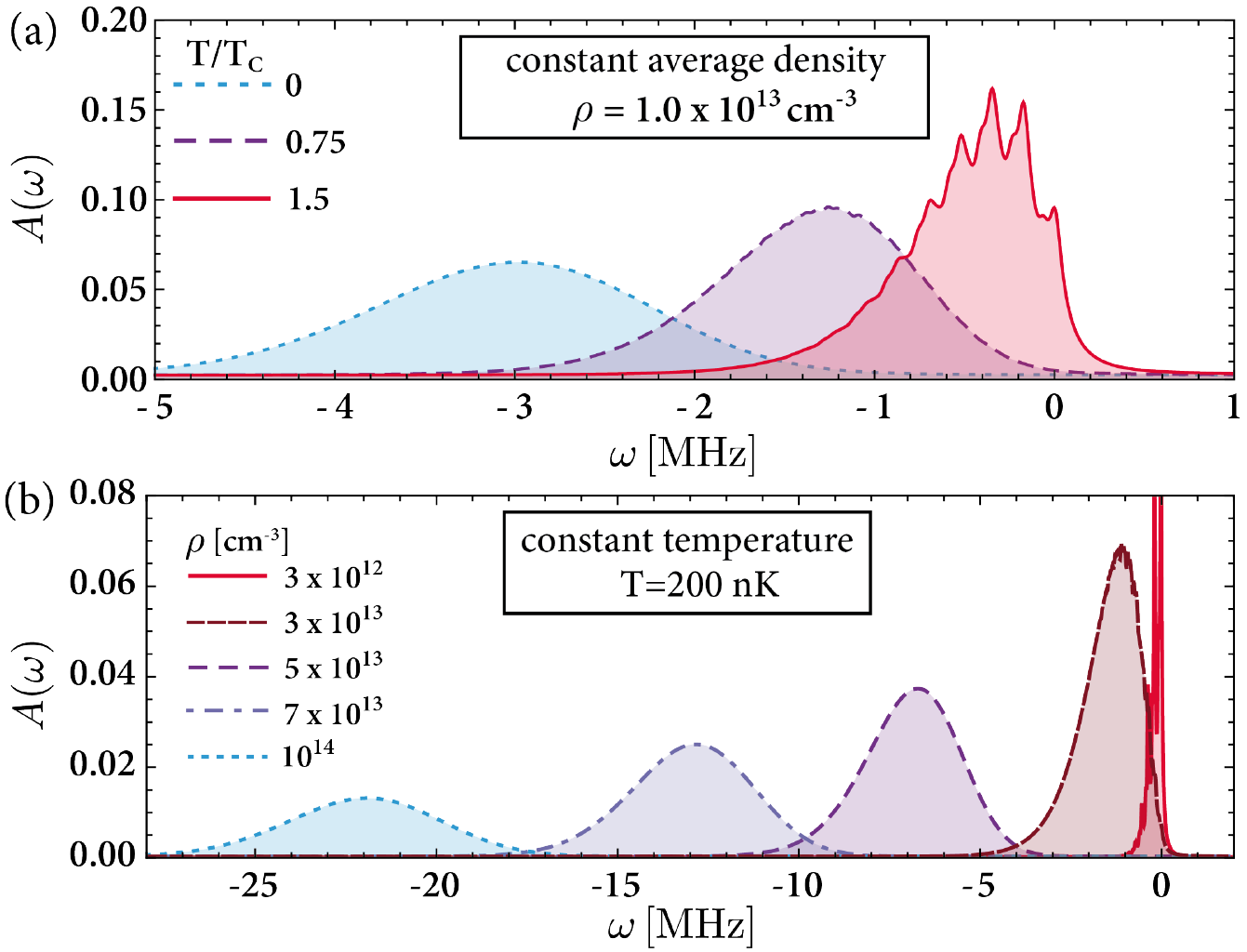}
        \caption{Temperature and density dependent absorption spectrum (excitation pulse length $70\mu$s).  (a) Spectral dependence on the temperature at fixed average density $\rho=10^{13}\,\text{cm}^{-3}$. (b)  Absorption spectrum at fixed temperature $T=200$nK for different average densities of the Bose gas. These spectra can be observed in experiments with crossed narrow-beam lasers \cite{nguyen2015,schlagmueller2015}.
        }
        \label{fig.TDep} 
\end{figure}

\textit{Finite temperature description.---} At finite temperatures, the gas is composed of $N_0$ condensed and $N'$ thermally depleted atoms. In order to account  for their quantum statistics, we make use of a novel bosonic functional  determinant approach (bFDA) which is based on earlier work in the context of mesoscopic physics for fermionic systems \cite{levitov1993,levitov1996}; see also \cite{klich_03} where bosonic systems are briefly discussed. 

Our method is not restricted to Rydberg systems and applies generally to impurities interacting with a bosonic environment. The FDA deals with many-body systems which are described within a grand-canonical ensemble. While this makes the method directly applicable to fermions with fixed particle number at zero temperature due to Pauli blocking, care is required for its application to atomic BEC: while  the atoms in the thermal cloud are still characterized by the density matrix $\hat \rho_{gc} = \exp{[-\beta (\hat H_0-\mu \hat N)]}/Z_G$  with partition function $Z_G$, the condensed fraction with fixed particle number cannot be described with the grand-canonical ensemble \cite{ziff1977}.  In our approach we take this into account by treating the thermal cloud and the condensate as separate subsystems. This approximation leads to the overlap 
\begin{equation}\label{fulloverlap}
S(t)=S_\text{BEC}(t)\times S_\text{th}(t),
\end{equation}
 where $S_\text{BEC}(t)$ refers to the time-dependent overlap  of the condensate and $S_\text{th}(t)$ to the contribution from the thermal cloud. Note, that the expression \eqref{fulloverlap} becomes exact both in the limit $T/T_c\ll1$ and $T/T_c\gg1$. 
 
In general, many-body traces as in Eq.~\eqref{overlap}, are difficult to evaluate due to the exponential size of Hilbert space. However, making use of the functional determinant approach, the thermal contribution can be expressed as \cite{klich_03} 
\begin{equation}\label{SThermal}
S_\text{th}(t)=\det\left[1+\hat n - \hat n e^{i\hat h_{0} t}e^{-i\hat h t}\right]^{-1}.
\end{equation}
Here $\hat n=1/[e^{\beta(\hat h-\mu) }-1]$ denotes the single particle occupation number operator, $\mu$ is the chemical potential and $\beta=1/k_BT$ with $k_B$ the Boltzmann constant. The operators $\hat h$, $\hat h_0$ are the  single-particle counterparts of $\hat H$ and $\hat H_0$. Eq.~\eqref{SThermal} allows for an efficiently transform of the many-body trace into a determinant in single-particle space, reducing computational complexity. We emphasize that eigenstates of $\hat h$ include all bound and scattering states in the presence of the Rydberg excitation. Thus  the bFDA is a non-perturbative approach that allows for including atomic potentials exactly.

In Fig.~\ref{fig.TDep}(a) we show the temperature dependence of the absorption spectrum at fixed average density. As  temperature is increased, the Gaussian profile at $T=0$ (dotted line) morphs in shape and detuning to an asymmetric profile with resolved molecular signatures of a dimer, trimer, etc. These molecular lines differ from the few-body spectra [cf.~Fig.~\ref{fig71s}] due to the dressing by fluctuations in the medium. 

In contrast to Fig.~\ref{fig.TDep}(a), in (b) the temperature is fixed and the average density is varied. At low densities (solid line) the gas is purely thermal and molecular lines are observed at low detuning (note the difference in frequency scale in Fig.~\ref{fig.TDep}). As the density is increased, and the  gas condenses, the spectrum moves from an asymmetric distribution to a Gaussian profile located at a detuning corresponding to hundreds of atoms bound within the orbit of the Rydberg impurity electron.

\begin{figure}[t]
        \centering
        \includegraphics[width=\linewidth]{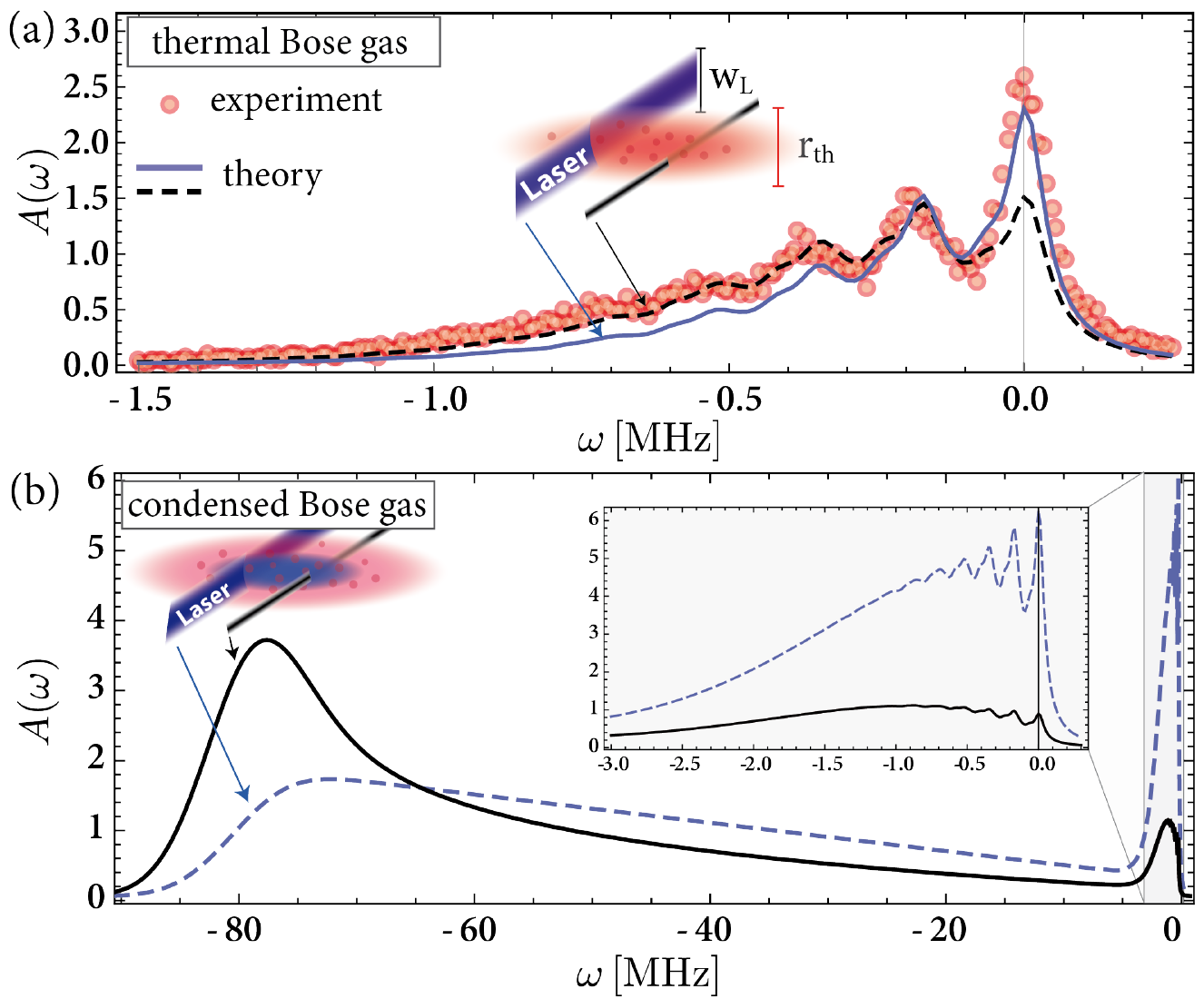}
        \caption{Density averaged absorption spectra for a Rb Rydberg excitation into the 71s state (excitation pulse length $30\mu$s). (a) Thermal gas at temperature $500\,\text{nK}$ and density in the center of the trap of $\rho_0=10^{13}\text{cm}^{-3}$. The results from bFDA are shown as lines for a laser profile illuminating the cloud over the full cylindrical radius $r_\text{th}$ (blue) and for a laser of small waist (dashed, black).  For illustration the data reported in by Gaj \textit{et al.} \cite{gaj2014,gajprivcom2015} is shown as red dots. (b) Density average absorption spectrum at temperature $T/T_c=0.87$ for a partially condensed gas of (zero temperature) peak density $\rho_p=5.5 \times 10^{14}\text{cm}^{-3}$. The response for two laser profiles is shown as in (a). The inset shows a zoom-in to the low-frequency regime.
        }
        \label{fig.LDA} 
\end{figure}

\textit{Experimental realization.---}
We have thus far only considered spectra at a well  defined density and temperature. However, in experiments \cite{gaj2014,nguyen2015,DeSalvo2015,schlagmueller2015}, the atomic clouds have an inhomogeneous density due to parabolic confinement and are probed by lasers of finite waist. In Fig.~\ref{fig.LDA} we present Rydberg spectra that take into account these density inhomogeneities  using a local density approximation (LDA). The analysis is done for a $^{87}\text{Rb}$ gas with realistic trap parameters, temperature, densities, and Rb-Rb scattering length \cite{sup}. In our calculations, we assume a Thomas-Fermi profile for the condensate while the thermal fraction is treated within the Hartree-Fock approximation \cite{Pitaevskii2003}.

In Fig.~\ref{fig.LDA}(a), we show the resulting averaged spectra for two laser profiles. The solid line represents a laser which excites the cloud along its full radial extent while the dashed line corresponds to a laser of a  narrow waist. The LDA leads to an  asymmetric line shape with long tail with clear signature of molecular oligomer excitations. For a laser of small waist, we predict a shift of spectral response to deeper detuning which can be tested in experiments. We emphasize that these first-principles calculations are performed without any adjustable parameters, except for the finite length of the laser pulse. For illustration we  compare our theoretical calculations to the molecular spectrum for $^{87}\text{Rb}$(71s) observed by Gaj \textit{et al.} \cite{gaj2014}. 

In Fig.~\ref{fig.LDA}(b), we present the density averaged absorption spectra for a gas below the BEC transition temperature $T/T_c<1$ \cite{sup}. The response predicted in Fig.~\ref{fig.LDA}(b)  results from the average over Gaussian distributions measured at different densities, and accounts fully for quantum statistics. While the response of the condensed atoms leads to a broad shoulder at large detuning, the thermal atoms contribute to a strong response at small detunings [inset in Fig.~\ref{fig.LDA}(b)]. The onset of LDA spectra at large detuning is temperature dependent and with increasing temperature, it will gradually move to small detunings. This theoretical prediction can be tested experimentally. 

As in Fig.~\ref{fig.LDA}(a), the line shape depends on the specifics of the spatial laser profile. In a recent experiment studying Rydberg impurity excitations in a $^{87}\text{Rb}$ BEC \cite{nguyen2015,schlagmueller2015}, such LDA profiles were interpreted using a classical treatment of atoms inside the Rydberg orbit, whose detunings were calculated from a combined s- and p-wave scattering of the Rydberg electron from the ground state atoms. Such a classical approach does not account for quantum many-body effects.

In an ultracold strontium gas \cite{DeSalvo2015}, p-wave scattering of the Rydberg electron leads to binding to the Sr atom, while in alkali metal atoms, including Rb, this scattering is resonant. Therefore, and for their higher densities, trapped Sr gases represent  a promising testing ground for probing the quantum many-body effects as well as the molecular shell structure.

\textit{Summary and Outlook.---} We presented a novel many-body formalism for analyzing non-equilibrium time evolution of Bose gases interacting with spatially extended impurities. We demonstrated that spectroscopy of Rydberg excitations allows for probing this dynamics in a new parameter regime where the impurity itself is extended over the entire size of the system.
From the analysis of the time-resolved overlaps of Rydberg atoms excited in an ultracold Bose gas we calculate the absorption spectra and find qualitative changes with temperature and density reflecting different regimes of dynamics.  At low densities we find the existence of a molecular shell structure and we predict the emergence of a metastable superpolaronic state at large densities and low  temperatures whose spectral signature is  a Gaussian distribution. 

Rydberg excitations in fermionic gases are another exciting direction for exploring impurity dynamics on mesoscopic scales, and the functional determinant approach is ideally suited to study such dynamics  \cite{knap2012,Schmidtprep,Cetinainprep}. For instance it is an open question whether Pauli blocking effects are observable  with a resulting  structure similar to the nuclear shell model. Theoretically our approach allows to efficiently calculate the real time evolution of occupation numbers, in situ density, and correlation functions. Those can be probed in future experiments aiming at  direct observation of shell structure, and the study of locally controlled heating of ultracold gases by Rydberg impurities. 

The coherence of Rydberg excitations is paramount for their application in quantum information processing \cite{saffman2010,gaj2014}. Our approach allows to investigate density dependent  dephasing for arbitrary dynamic protocols and can be used to find optimal protocols for preparing interesting quantum states. Theoretical predictions of our analysis can be tested further either with spectroscopic methods or with interferometric probes \cite{Cetinainprep}.

\textit{Acknowledgements.---} 
R.~S. and H.~R.~S. were supported by the NSF through a grant for the Institute for Theoretical Atomic, Molecular, and Optical Physics at Harvard University and the Smithsonian Astrophysical Observatory. E.~D. acknowledges support from Harvard-MIT CUA, NSF Grant No. DMR-1308435, AFOSR Quantum Simulation MURI, the ARO-MURI on Atomtronics, and support from  Dr.~Max R\"ossler, the Walter Haefner Foundation and the ETH Foundation.  We are thankful to A.~Gaj and T.~Pfau for providing data.

\end{document}


\title{Supplemental Material on \\ 
A mesoscopic Rydberg impurity in an atomic quantum gas}

\author{Richard Schmidt}
\affiliation{ITAMP, Harvard-Smithsonian Center for Astrophysics, Cambridge, MA 02138, USA}
\affiliation{Department of Physics, Harvard University, Cambridge MA 02138, USA}
\author{ H. R. Sadeghpour}
\affiliation{ITAMP, Harvard-Smithsonian Center for Astrophysics, Cambridge, MA 02138, USA}
\author{E. Demler}
\affiliation{Department of Physics, Harvard University, Cambridge MA 02138, USA}

\maketitle

\textbf{Single-particle solutions.---} In Eq.~(7), the matrix elements of the single-particle operator [we work in units $\hbar=1$]
\begin{equation}\label{SMatrix}
\hat C\equiv1+\hat n - \hat n\, e^{i\hat h_{0} t}e^{-i\hat h t},
\end{equation}
are calculated in the non-interacting single particle basis for an atom trapped in a spherical box of radius $R=2\times 10^5\,a_0$. We insert in Eq.~\eqref{SMatrix} a complete set of eigenstates of the interacting single particle Hamiltonian $\hat h$, determined from the solutions of the Schr\"odinger equation
\begin{equation}\label{SEquation}
\left[-\frac{\Delta}{2 m_B}+V_\text{Ryd}(\vecr)\right]\psi(\vecr)=E\psi(\vecr),
\end{equation}
where  $m_B$ is the mass of the medium bosons (denoted by $m$ in the main text). The spherically symmetric Rydberg molecular potential is given in Eq.~(1).  We express the eigenfunctions of Eq.~\eqref{SEquation} as $\psi(\vecr)=\frac{u_{nl}(r)}{r}Y_{lm}(\Omega)$ where $n$ labels the `radial' quantum number, $l$ the angular momentum and $m$ its projection. The radial wavefunctions $u_{nl}(r)$ are $m$ independent and are obtained as solutions to the equation
\begin{equation}\label{radialEq}
\epsilon_{nl} u_{nl}(r) = \left[-\frac{1}{2m_B}\frac{\partial^2}{\partial r^2}+\frac{l(l+1)}{2m_Br^2}+V_\text{Ryd}(r)\right]u_{nl}(r).
\end{equation}
Note, in our model the Rydberg impurity is assumed to have an infinite mass. We rescale the electron scattering length $a_e$ in Eq.~\eqref{radialEq} and Eq.~(2)  by a factor of $0.85$, so that the observed vibrational energies are reproduced.

\begin{figure}[b]
        \centering
        \includegraphics[width=\linewidth]{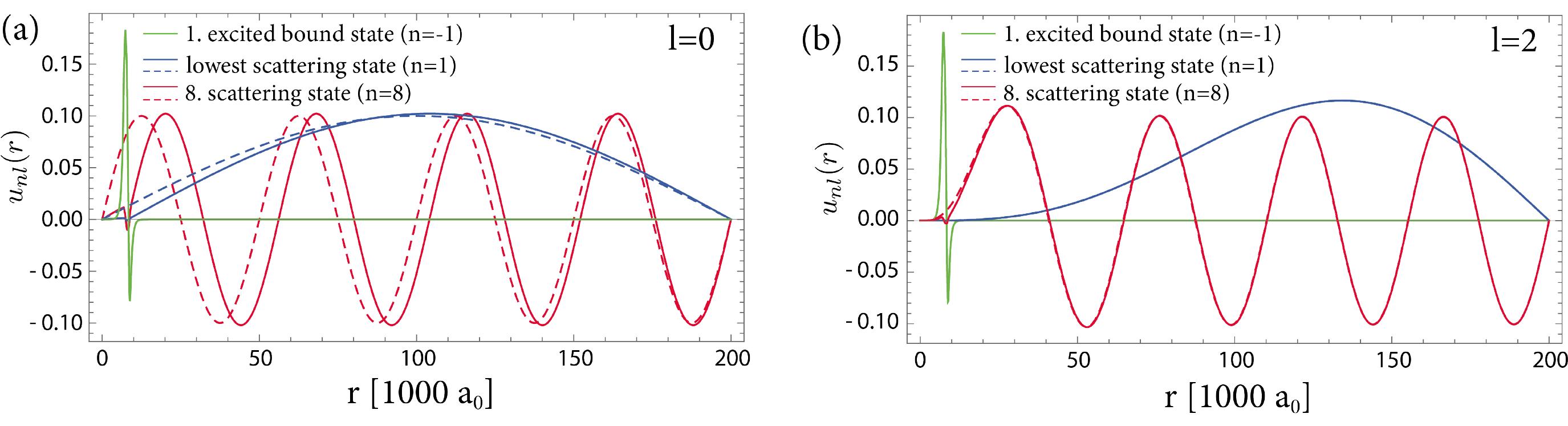}
        \caption{Solutions of the radial Schr\"odinger equation. We compare non-interacting solutions (dashed) with the single particle solutions in presence of the Rydberg potential resulting from an Rb(71s) excitation. (a) Angular momentum $l=0$. We show the solutions for radial quantum numbers $n=1,8$ (blue, red) as well as the first excited bound state (green) which is also shown in the inset of Fig.~2(a). (b) Angular momentum $l=2$. We show the solutions for radial quantum numbers $n=1,8$ (blue, red) as well as the first excited bound state (green). 
        }
        \label{fig.sup} 
\end{figure}

The radial Schr\"odinger equation Eq.~\eqref{radialEq} is  solved numerically for the first 300 scattering wave functions. We consider states with $l\leq 8$ and have verified that higher angular momenta do not contribute significantly  at the temperatures of interest. In Fig.~\ref{fig.sup} we show the resulting wave functions for various radial quantum numbers $n$ and angular momenta $l$. We evaluate the matrix elements in Eq.~\eqref{SMatrix} for a non-interacting basis set of size $\sim2400$ by invoking the conservation of the angular momentum. Finally we calculate its determinant which yields the overlap function $S(t)$.\\

\textbf{Emergence of the Gaussian spectrum.---} 
An explicit analytical expression for the spectral decomposition of the zero temperature overlap in Eq.~(4) can most conveniently be derived directly from Eq.~(3) by a basis change from the single particle orbitals $\ket{i}$, defined by $\hat h_0 \ket{i}=\epsilon_i \ket{i}$, and labeling the creation and annihilation operators $\bed_i$ and $\be_i$  to the interacting basis spanned by the states $\ket{\alpha}$, defined by $\hat h \ket{\alpha}=\epsilon_\alpha \ket{\alpha}$. Following standard contractions rules \cite{fetter2003} and making use of the multinomial identity
\begin{eqnarray}
\left(x_1+\ldots + x_M\right)^N&=& \sum_{\sum p_i=N}\binom{N}{p_1, \ldots, p_M}\,\,x_1^{p_1}\ldots x_M^{p_M}
\end{eqnarray}
one finds (setting the energy of the lowest non-interacting state $\ket{s}$ to zero) \cite{Schmidtprep}

\begin{eqnarray}\label{SpecAna}
A(\omega)=N! \sum_{\Sigma p_i = N}\frac{|\braket{\alpha_1}{s}|^{2 p_1}\cdot \ldots \cdot|\braket{\alpha_M}{s}|^{2 p_M}\cdot \ldots}{p_1!\ldots p_M! \ldots} \,\,\,\delta(-\omega+p_1 \omega_1+\ldots +p_M \omega_M+\ldots)
\end{eqnarray}
where the sum goes over all permutations to distribute $N$ particles into the infinitely many interacting single-particle states $\ket{\alpha_i}$ in presence of the Rydberg impurity. The state $\ket{0}\equiv\ket{s}$ is macroscopically occupied in the BEC state, $\ket{\Psi_0}=1/\sqrt{N!} (\be^\dagger_0)^N\ket{\text{vac}}$. The expression \eqref{SpecAna} describes the stochastic distribution of atoms out of the lowest trap state $\ket{s}$ into interacting single-particle states $\ket{\alpha_i}$ in presence of the Rydberg impurity. Each of those processes yields a delta contributions to the spectrum (in the thermodynamic limit the discrete spectral distribution (or spectral function) becomes continuous) which represent the eigenenergies of the many-body Hamiltonian. Each of those eigenstates is weighted by the prefactor in Eq.~\eqref{SpecAna} which is determined by a combination of Frank Condon factors $|\braket{\alpha_i}{s}|^{2}$. 
 
Since the Frank-Condon factors fulfill the relation $\sum_i |\braket{\alpha_i}{s}|^{2}=1$, the prefactor in front of the delta function represents a multi-nominal distribution. Furthermore, since the non-interacting state $\ket{s}$ has appreciable overlap only with a small number of interacting single particle states $\ket{\alpha_i}$,  the sum in Eq.~\eqref{SpecAna} has only a few contributions. Hence, for large enough particle number $N$, the multinomial distribution approaches a Gaussian. In the case that only two interacting single particle states, say $\ket{\alpha_1}\equiv\ket{B}$ and $\ket{\alpha_2}\equiv\ket{\tilde s}$ have finite overlap with $\ket{s}$, the expression \eqref{SpecAna} reduces to Eq.~(5) (for $\omega_{\tilde s}\to 0$).
 
Eq.~\eqref{SpecAna} is the analytical Fourier transform of Eq.~(4). However, while Eq.~\eqref{SpecAna} is  instructive for a qualitative assessment of spectral properties, a direct calculation of an absorption spectrum using Eq.~\eqref{SpecAna} is  difficult due to the exponential growth of the number of states involved. Hence for the calculation of spectra we use the numerical Fourier transform of the overlap function $S(t)$. \\

\textbf{Details on the local density approximation.---}  For the calculation of the absorption spectra shown in Fig.~4 we employ a local density approximation for a gas of \Rb atoms in a harmonic trap with cylindrical shape with frequencies $\omega_r=200\,\text{Hz}$ and $\omega_y=15\,\text{Hz}$ \cite{nguyen2015}. For Fig.~4(a), we perform the calculations for a thermal gas at $T=500$nK with a trap peak density of $\rho_p=10^{13}\,\text{cm}^{-3}$. The calculation of the density profile is done in a Hartree-Fock approximation \cite{Pitaevskii2003} which takes into account the effective repulsion between the bosons due to the positive s-wave inter-boson scattering length $a_\text{Rb}=100\,a_0$ \cite{gaj2014}. 

The LDA spectrum is obtained assuming that the laser illuminates the central region with respect to the elongated axis of the cloud [cf.~the illustration in Fig.~4(a)]. We show the results for two choices of the radial waist $w_L$ of the laser. First, for a laser with waist  which illuminates the cloud along an one dimensional line, $w_L/r_\text{th}\to 0$, with $r_\text{th}$ denoting the radius where the thermal cloud density falls off to a density of $0.01 \,\rho_p$ [black dashed line in Fig.~4(a), and blue solid line  in Fig.~4(b)], and, second, for a laser which illuminates the cloud along the whole radial direction [black solid line in Fig.~4(a), and blue dashed line  in Fig.~4(b)]. 

In Fig.~4(b), we assume a gas of particle number $N=1.5\times 10^6$ atoms, which at zero temperature leads to a Thomas-Fermi profile with peak density $5.5\times 10^{14}\text{cm}^{-3}$. The trap frequencies are the same as for Fig.~4(a) \cite{nguyen2015}. The spectra shown  in Fig.~4(b) are calculated for a temperature $T/T_c=0.87$. Here a fraction of the Bose gas is  condensed and the thermal cloud is partially expelled from the center of the cloud due to the inter-boson repulsion accounted for in the Hartree-Fock approximation \cite{Pitaevskii2003}. To obtain the averaged spectra, the sampling is performed  at $\sim100$ densities along the radial axis and we  determine the local condensate and thermal densities which become the input in the functional determinant calculation of $S(t)$ in Eq.~(6).


%
